\documentclass[conference]{IEEEtran}
\usepackage{lipsum}
\usepackage{stfloats}
\usepackage{cite}
\usepackage{graphicx}
\usepackage{amsmath}
\usepackage{amsfonts}
\usepackage{balance}
\usepackage{fancyhdr}
\usepackage{graphicx,times,cite,amssymb,epsfig,amsthm,color}
\usepackage[normalem]{ulem}
\usepackage{xcolor}
\usepackage{diagbox}

\newlength{\figwidth}
\renewcommand{\baselinestretch}{1}    
\normalsize

\usepackage{graphicx}
\ifCLASSINFOpdf
\else
\fi
\usepackage{amsmath}
\DeclareMathOperator*{\argmax}{argmax}
\DeclareMathOperator*{\argmin}{argmin}

\usepackage[cmintegrals]{newtxmath}
\usepackage{algorithm}
\usepackage{algorithmic}

\usepackage{array}

\usepackage{stfloats}
\usepackage{url}

\usepackage{verbatim}

\makeatletter
\newcommand*{\rom}[1]{\expandafter\@slowromancap\romannumeral #1@}
\makeatother

\newcommand{\numfreqs}{M}
\newcommand{\numslots}{L}

\newcommand{\pri}{T}

\newcommand{\freqsep}{\Delta f}

\newcommand{\pulset}{p}

\newcommand{\numantas}{N}
\newcommand{\re}{\text{Re}}

\begin{document}
%
\title{Can FSK Be Optimised for Integrated Sensing and Communications?}
%
%
%

\author{Tian~Han,~\IEEEmembership{Student~Member,~IEEE,}
        Peter~J~Smith,~\IEEEmembership{Fellow,~IEEE,}
        Urbashi~Mitra,~\IEEEmembership{Fellow,~IEEE,}
        Jamie~S~Evans,~\IEEEmembership{,Senior~Member,~IEEE,}
        and~Rajitha~Senanayake~\IEEEmembership{Member,~IEEE}
        }
\thanks{This work was supported by the Australian Research Council Discovery Project under Grant DP200103625, Army Research Office under Grant W911NF1910269, Swedish Research Council under Grant 2018-04359, National Science Foundation under Grants RINGS-2148313, CCF-2200221 and CIF-2311653, and Office of Naval Research under Grant N00014-15-1-2550.}

\maketitle

\begin{abstract}
Motivated by the ideal peak-to-average-power ratio and radar sensing capability of traditional frequency-coded radar waveforms, this paper considers the frequency shift keying (FSK) based waveform for joint communications and radar (JCR). An analysis of the probability distributions of its ambiguity function (AF) sidelobe levels (SLs) and peak sidelobe level (PSL) is conducted to study the radar sensing capability of random FSK. Numerical results show that the independent frequency modulation introduces uncontrollable AF PSLs. In order to address this problem, the initial phases of waveform sub-pulses are designed by solving a min-max optimisation problem. Numerical results indicate that the optimisation-based phase design can effectively reduce the AF PSL to a level close to well-designed radar waveforms while having no impact on the data rate and the receiver complexity. For large numbers of waveform sub-pulses and modulation orders, the impact on the error probability is also insignificant.

\end{abstract}


%
\IEEEpeerreviewmaketitle

\section{Introduction}  \label{sec:intro}

The ongoing roadmapping of next-generation wireless networks has considered applications and services including intelligent vehicular networks, the Internet of Things and smart living, in which both sensing and communications play critical roles \cite{liu22abb, Wei23, kaushik23}. Therefore, integrated sensing and communications (ISAC) and joint communications and radar (JCR), as one mainstream in ISAC, have attracted extensive research interest. Primarily motivated by the fact that both functions are moving towards the millimetre wave (mmWave) frequency band, JCR has been considered a promising paradigm that can contribute to reusing the precious spectrum resource and reducing the hardware device size and power consumption. 

\subsection{Related Work}
An important aspect of integrating communications and radar sensing in a dual-function system is the design of a waveform which has the capability of transmitting data and performing radar sensing simultaneously. Such waveforms are generally designed from three different perspectives, namely, communications-centric, radar-centric and joint communication-radar design \cite{Aliu2022abb}.

In communication-centric methods, the focus is on integrating radar sensing into existing communications systems. For example, IEEE 802.11ad-based waveforms with well-designed preambles, such as the Golay complementary sequences for stationary target sensing \cite{Grossi20} or the Prouhet-Thue-Norse sequences \cite{Pezeshki08} to combat Doppler effects \cite{Duggal20abb}, have been considered as suitable JCR candidates. 
This design method results in a comparable data rate to pure communications systems. However, the limited time duration of the preamble introduces a fundamental constraint on the radar sensing performance of those preamble design methods. 

From a similar communications-centric perspective, there has also been strong interest in the use of orthogonal frequency division multiplexing (OFDM) based waveforms for JCR \cite{Sturm11}. Different resource allocation schemes across OFDM frequency bins are examined to optimise the performance. The optimisation can be maximising the mutual information (MI) between the received samples and the target scattering \cite{bicua2016generalized}, or maximising the communications capacity lower bound while keeping the similarity to a radar-optimal waveform \cite{keskin2021limited}. When combined with triangular frequency modulation, improved Doppler performance can be achieved \cite{Wang22}. Despite the benefits of adapting a legacy signalling scheme and the associated high data rates, OFDM suffers from high peak-to-average power ratios (PAPRs) and results in non-linear distortion at the radio frequency front-end power amplifier \cite{Huang22}. To this end, \cite{Huang22} and \cite{Piyush23} use only a portion of the OFDM subcarriers for data transmission while the remaining subcarriers are designed to minimise the waveform PAPR using majorisation-minimisation techniques. Lower PAPR values are achieved by sacrificing the data rate.

In contrast, radar-centric strategies focus on embedding information into traditional radar waveforms. In \cite{Blunt10}, the traditional linear frequency modulated (LFM) waveform is considered for JCR. However, the data rate is non-ideal since the communications symbol rate corresponds to the chirp rate only. Stepped frequency radar waveforms have a deterministic control of the time-bandwidth product of the waveform and a unit baseband PAPR, resulting in ideal radar estimation performance \cite{richards14,Levanon2004}. Motivated by these good properties, \cite{Senanayake22TWCabb} embeds data into the stepped frequency waveform via permutations of the frequency sequence, which allows data transmission while keeping good delay and Doppler shift estimation capabilities on average. The communication error rate or the radar ambiguity function (AF) peak sidelobe level (PSL) can be further improved by using subsets of these permutations, though it leads to a data rate reduction \cite{Dayarathna23}. Similar to LFM-based designs, the data rate of stepped frequency-based designs is usually non-ideal. To increase the data rate, \cite{Han2023_freqpermPSK} incorporates phase shift keying (PSK) into waveform sub-pulses while having little impact on the radar performance. 

Data transmission using carrier agile phased array radar (CAESAR) can be achieved via index modulation (IM) by embedding communication signals using combinations or permutations of radar signal parameters such as frequency, space, bandwidths, and/or polarisations \cite{Ma20joint, huang20, Temiz23}, resulting in high data rates with robust radar performance. IM can also be applied to frequency-hopping (FH) multi-input multi-output (MIMO) radar systems by using frequency codes among fast-time sub-pulses transmitted from different antennas \cite{Wu22overviewabb}. The data rate can be improved further using hybrid modulation strategies, which combine frequency codes with quadrature amplitude modulation (QAM) or PSK \cite{Baxter22abb}. The use of PSK can also reduce AF sidelobe levels (SLs) \cite{Eedara18}. Nevertheless, high computational complexity is usually unavoidable in the optimal demodulation design of IM-based systems, leading to difficulties in practical implementation \cite{zhang2021abb}.

Unlike communications-centric or radar-centric design methods which rely on existing waveforms, the joint design methods aim at designing JCR waveforms from the ground up \cite{liu22abb}. This type of method usually provides extra flexibility to trade-off between various communications and radar performance metrics and allows for improving both aspects simultaneously. These works usually formulate optimisation problems which couple radar and communication performance including multi-user interference, power, rate-distortion and radar sensing mean squared error \cite{Liu18, Dong23}. A common challenge is the non-convexity of the objective function which leads to approximate and sub-optimal solutions.

There also have been waveform design problems aiming to solve the challenges of co-existence. More specifically, in a situation where separate radar and communications transmitters send different signals simultaneously while sharing the same frequency spectrum, the waveform design focuses on mitigating interference in the system while satisfying various radar and communications performance metrics \cite{zheng2019radar}. The performance metrics include the received signal-to-interference-and-noise-ratio (SINR), the MI between the radar target and the received signal \cite{huang2015radar}, the communications link capacity, and/or the radar estimation rate \cite{chiriyath2017radar}. While most co-existence designs consider information theoretical metrics for sensing, a more straightforward relationship between these metrics and the actual estimation performance remains an open problem.

\subsection{Motivation and Contributions}



In this paper, we consider a communications-centric design by considering waveforms whose sub-pulses are modulated using $\numfreqs$-ary FSK. Although it is a communications-centric design, there exist some similarities to the stepped frequency radar waveforms \cite{richards14} since both have one single carrier frequency in each waveform sub-pulse. In addition, both waveforms share the ideal property of a unit baseband PAPR, which is beneficial for both radar and communications. A high PAPR value can limit the practical performance of a waveform since non-linear devices such as the power amplifier at the radio frequency front end are very sensitive to high PAPR values \cite{Huang22}. The distortion introduced by the non-linearity of those devices can degrade both communications error probability and radar estimation performance \cite{Liu23_70years}. 

The use of FSK for JCR has attracted an increasing amount of interest since the independent data modulation from sub-pulse to sub-pulse leads to a data rate comparable with traditional communications schemes \cite{Han23_globecom}. However, the radar estimation performance, in terms of both time-bandwidth properties and the AF SLs, is not directly controlled. A compact theoretical analysis of its impact on the time-bandwidth performance has been conducted in \cite{Han23_globecom}. 
Nevertheless, the analysis of its AF SL and PSL remains an open problem. 

Although the AF SL and PSL are widely used radar performance measures, it is difficult to quantify them analytically due to the complicated expression of the AF. This is even more difficult in JCR systems since this challenging problem has to be solved for multiple waveforms. In this work, we simplify the problem by following the idea for designing Costas frequency-coded radar waveforms \cite{Levanon2004}, i.e., focusing on the AF SLs at grid points of the delay-Doppler plane. To the best of our knowledge, this work is the first to analyse and suppress the AF SLs for FSK waveforms. 
The main contributions of this paper are as follows.
\begin{itemize}
\item We derive the AF expressions for the FSK-based waveform, based on which the exact expressions for the probability distributions of the grid point SLs are derived. We provide an accurate approximation of the AF PSL distribution by using the fact that the correlation between any two SLs is zero or low.
\item In order to deal with the uncontrolled AF PSL performance introduced by independent FSK, we propose a PSL suppression method which designs the initial phase of each sub-pulse by solving a min-max optimisation problem whose objective function is the AF PSL. 
\item 
The impact of the PSL suppression on the communications performance is assessed.
\item Numerical results show that the proposed method leads to a PSL close to the well-designed Costas coded waveform \cite{Levanon2004}, without affecting the data rate and the unit baseband PAPR property of FSK. 
\item Numerical results also show that the proposed waveform achieves less than half the AF PSL and a lower PAPR on average compared to a recent low PAPR OFDM JCR waveform design \cite{Piyush23} with the same data rate.
\end{itemize}
The remainder of this paper is organised as follows. The FSK signal model and the performance metrics are introduced in Section \ref{sec:prob}. The AF SL and PSL distributions are analysed in Section \ref{sec:prob}. In Section \ref{sec:AF_sll}, the optimisation-based AF PSL suppression is proposed and its impact on communications performance is discussed. Numerical examples are provided in Section \ref{sec:num_FSK_global} to support our theoretical analysis and verify the effectiveness of the PSL suppression method.



\section{Problem Formulation} \label{sec:prob}


\subsection{Signal Model}
Consider a joint communications and radar system, in which the transmitter sends an $\numfreqs$-ary FSK-based pulsed waveform whose complex envelope can be expressed as
\begin{align}
s(t) = \sqrt{\frac{1}{\numslots \pri}} \sum_{l=0}^{\numslots-1}\pulset \left(t- l\pri \right) \exp(j\omega_l(t-l\pri)), \label{eq:wf_FSK}
\end{align}
where $\numslots$ is the number of sub-pulses, $\pri$ is the sub-pulse repetition interval and $\pulset(t)$ is the rectangular pulse shaping function which can be expressed as 
\begin{align} \label{eq:rect_pulse}
\pulset(t) = \left\{
\begin{array}{ll}1~~~~~~~~&0 \leq t \leq \pri \\
0~~~~~~~~&\text{otherwise.} 
\end{array}\right.
\end{align}
The amplitude, $\sqrt{1/\numslots \pri}$, is chosen such that $s(t)$ contains unit energy, i.e., $\int_{-\infty}^{\infty}|s(t)|^2dt = 1$. The frequency modulated onto the $l$-th sub-pulse, $\omega_l$, satisfies $\omega_l \in \{0, 2\pi \freqsep, \cdots, 2\pi(\numfreqs-1) \freqsep\}$, where $\freqsep$ is the separation between two neighbouring frequencies. In order to keep orthogonality between two sub-pulses with different centre frequencies, $\freqsep$ should satisfy $\freqsep = i/\pri$, where $i$ is a non-zero integer.

\subsection{Radar Performance Measures}
The AF is an important analytical tool for designing radar waveforms and analysing radar-sensing performance \cite{richards14}. More specifically, it describes the output of a matched filter when the received signal is time delayed by $\tau$ and Doppler shifted by $\omega$. The complex AF of a waveform $s(t)$ can be expressed as \cite[eq.(4.30)]{richards14}
\begin{align} \label{eq:cAF}
    \hat{A}(\tau, \omega) = \int_{-\infty}^{\infty} s(t) s^*(t-\tau)e^{-j\omega t} d t.
\end{align}
The AF is defined as the magnitude of the complex AF which is given by
\begin{align} \label{eq:AF}
A(\tau, \omega) = \left| \int_{-\infty}^{\infty} s(t) s^*(t-\tau)e^{-j\omega t} d t \right|. 
\end{align}
Generally speaking, the AF performance can be separated into local accuracy and global accuracy \cite[Chapater 10]{vantrees01}. The local accuracy is characterised by the AF performance around the origin of the delay-Doppler plane, while the global accuracy is characterised by its SLs. 
In this paper, we focus on analysing the global accuracy of the FSK waveform based on the AF SLs. As will be discussed in the following paragraphs, this is a critical and difficult problem for radar.

One single measure that is often used to summarise the overall AF SL performance is the PSL. The PSL is the largest SL of an AF in the delay-Doppler plane. It can be used to measure the performance of a radar waveform against clutter in the worst-case scenario, i.e., the clutter presents at the delay-Doppler values corresponding to the AF PSL. Clutter, such as ground or sea surface, usually has a large radar cross-section, meaning that the signal power reflected by clutter is usually large. If the clutter enters via the delay-Doppler values corresponding to the PSL, at the radar receiver output it can easily beat the target, resulting in incorrect estimation. Therefore, the AF PSL is a crucial metric to analyse. 

The exact definition of the AF PSL depends on the definition of the AF SL. One common definition of an SL is the AF at a local maximum. However, its quantitative analysis is extremely difficult since it requires taking the derivative of the complicated AF and setting it to zero. Closed-form solutions for this calculation are almost impossible to calculate analytically. This problem is even more challenging for JCR waveforms, as this process must be performed for multiple AFs resulting from different transmitted waveforms. Motivated by the design method of the Costas-coded radar waveform \cite[Chapter 5]{Levanon2004}, in this work we simplify the problem by focusing on the AF values at the grid points, i.e., $\tau = k \pri$ and $\omega = 2\pi r \freqsep$, where $k$ and $r$ are integers. Similar simplification has been considered in \cite{Peer22} to analyse the AF SLs for random phase-coded JCR waveforms. This allows us to capture the impact of the frequency modulation and study the statistical performance of the SL. The grid point PSL is then the maximum of all grid point SLs. Since fewer delay-Doppler values are considered compared to the local maxima PSL, the grid point PSL is a lower bound on the local maxima PSL. We will show in this work that the grid point PSL is a useful lower bound and evaluate the statistical performance of the grid point PSL.

\section{Ambiguity Function Sidelobes}  \label{sec:AF_sll}
\subsection{Ambiguity Function} 
In order to study the sidelobe performance, we first derive the expression of the AF. Substituting \eqref{eq:wf_FSK} into \eqref{eq:cAF}, the complex AF of the FSK-based waveform can be expressed as
\begin{align}\label{eq:cAF_FSK1}
\hat{A}(\tau, \omega) = & \frac{1}{\numslots \pri}  \sum_{l = 0}^{\numslots - 1}\sum_{n = 0}^{\numslots - 1} \int_{-\infty}^{\infty} \pulset(t-lT) \pulset^*(t-nT-\tau) \notag\\ &\times e^{j (\omega t + \omega_l(t-l\pri) - \omega_n(t-\tau-n\pri))} d t.
\end{align}
We note that the integral, i.e., $\int_{-\infty}^{\infty} \pulset(t-lT)\pulset^*(t-nT-\tau) e^{-j (\omega-\omega_n+\omega_l)(t-l\pri)} d t$ in \eqref{eq:cAF_FSK1} can be expressed as $\hat{A}_p(\tau+(n-l)\pri, \omega-\omega_n+\omega_l)$, where $\hat{A}_p(\tau, \omega)$ is the complex AF of $\pulset(t)$ which can be written as
\begin{align}\label{eq:cAF_rect}
\hat{A}_p(\tau, \omega) = \left\{
\begin{array}{ll} e^{j\omega \tau}\left(\frac{e^{j\omega \pri} - e^{-j\omega \tau}}{j\omega}\right)~~~~~~~~-\pri \leq \tau < 0\\
\frac{e^{j\omega \pri} - e^{j\omega \tau}}{j\omega}~~~~~~~~~~~~~~~~~~~~0 \leq \tau \leq \pri\\
0~~~~~~~~~~~~~~~~~~~~~~~~~~~~~~~\text{otherwise}.
\end{array}\right.
\end{align}
Substituting $\hat{A}_p(\tau+(n-l)\pri, \omega-\omega_n+\omega_l)$ into \eqref{eq:cAF_FSK1} and using some straightforward mathematical manipulations, we obtain
\begin{align}\label{eq:cAF_FSK}
\hat{A}(\tau, \omega) = \frac{1}{\numslots \pri}  \sum_{l = 0}^{\numslots - 1}\sum_{n = 0}^{\numslots - 1} &\hat{A}_p(\tau + nT - lT, \omega - \omega_n + \omega_l) \notag\\ &\times e^{j (\omega l \pri + \omega_n((n - l)\pri+ \tau) )}.
\end{align}

The AF of the waveform by obtained by taking the magnitude of $\hat{A}(\tau, \omega)$, i.e.,
\begin{align}\label{eq:AF_FSK}
A(\tau, \omega) = \Bigg|\frac{1}{\numslots \pri}  \sum_{l = 0}^{\numslots - 1}\sum_{n = 0}^{\numslots - 1} &\hat{A}_p(\tau + n\pri - l\pri, \omega - \omega_n + \omega_l) \notag\\ &\times e^{j (\omega l \pri + \omega_n((n - l)\pri+ \tau) )}\Bigg|.
\end{align}

The zero-Doppler cut of the AF, which describes the matched filter output when there is no Doppler shift, can be derived by letting $\omega = 0$ in \eqref{eq:AF}, resulting in
\begin{align}\label{eq:0Dop_FSK}
A(\tau, 0) = \Bigg|\frac{1}{\numslots \pri}  \sum_{l = 0}^{\numslots - 1}\sum_{n = 0}^{\numslots - 1} &\hat{A}_p(\tau + n\pri - l\pri, \omega_l - \omega_n) \notag\\ &\times e^{j \omega_n((n - l)\pri+ \tau) }\Bigg|.
\end{align}
Similarly, the matched filter output versus the Doppler shift $\omega$ when $\tau = 0$ is given by the zero-delay cut of the AF, which can be expressed as
\begin{align}\label{eq:0del_FSK1}
A(0, \omega) = \Bigg|\frac{1}{\numslots \pri}  \sum_{l = 0}^{\numslots - 1}\sum_{n = 0}^{\numslots - 1} &\hat{A}_p((n - l)\pri, \omega - \omega_n + \omega_l) \notag\\ &\times e^{j (\omega l \pri + \omega_n((n - l)\pri) )}\Bigg|.
\end{align}
Since $\hat{A}_p(\tau, \omega)$ is non-zero only when $|\tau| < \pri$, $\hat{A}_p((n - l)\pri, \omega - \omega_n + \omega_l)$ is non-zero only when $n=l$. Therefore, using straightforward mathematical manipulations, the zero-delay cut \eqref{eq:0del_FSK1} can be simplified as
\begin{align}\label{eq:0del_FSK}
A(0, \omega) = \Bigg|\frac{1}{\numslots \pri}  \sum_{l = 0}^{\numslots - 1}\hat{A}_p(0, \omega)  e^{j \omega l\pri}\Bigg|.
\end{align}
The zero-Doppler cut and the zero-delay cut are used to describe the delay estimation accuracy and the Doppler shift estimation accuracy, respectively, when the other parameter is known. We note from \eqref{eq:0del_FSK} that the AF zero-delay cut is independent of the frequency modulation. 
However, both \eqref{eq:0Dop_FSK} and \eqref{eq:AF_FSK} show that it is difficult to analyse the impact of frequency change on the SLs and the PSLs at the other parts of the delay-Doppler plane. In the following section, we study this problem by focusing on the statistical properties of the grid point SLs and PSLs. 

\subsection{AF Sidelobes at Grid Points} 
Due to the symmetry of the AF, i.e., $A(\tau, \omega) = A(-\tau, -\omega)$ \cite[Chapter 4]{richards14}, we focus on the first and the fourth quadrants of the delay-Doppler plane. Substituting $\tau$ with $k\pri$ and $\omega$ with $2\pi r \freqsep$ in \eqref{eq:0Dop_FSK} where $(k,r) \in \mathcal{D}$ with $\mathcal{D} = \{(k,r)| k \in \{0, \cdots, \numslots - 1\}, r \in \{-(\numfreqs-1), \cdots, \numfreqs - 1\}\} \setminus \{(k,r)| k \leq 0, r = 0 \}$ and defining $\Tilde{A}(k,r) = A(k \pri, 2\pi r \freqsep)$, the grid point SL can be expressed as
\begin{align}\label{eq:AF_FSK_grid1}
\Tilde{A}(k,r) = \Bigg|\frac{1}{\numslots \pri}  \sum_{l = 0}^{\numslots - 1}\sum_{n = 0}^{\numslots - 1} & \hat{A}_p((k-l+n)\pri, 2\pi r \freqsep - (\omega_n - \omega_l))\notag\\ 
&\times e^{j (2\pi r l \freqsep \pri + \omega_n(k-l+n)\pri)}\Bigg|.
\end{align}
Since $\hat{A}_p(\tau, \omega)$ is non-zero only when $|\tau| < \pri$, $\hat{A}_p((k-l+n)\pri, 2\pi r \freqsep - (\omega_n - \omega_l))$ is non-zero only when $k=l-n$. In addition, since $\freqsep \pri$ is an integer, $e^{j (2\pi r l \freqsep \pri + \omega_n(k-l+n)\pri)}$ is always $1$. Thus, \eqref{eq:AF_FSK_grid1} can be expressed as
\begin{align} \label{eq:AF_FSK_grid2}
\Tilde{A}(k,r) = \Bigg|\frac{1}{\numslots \pri}  \sum_{l = k}^{\numslots - 1} \hat{A}_p(0, 2\pi r \freqsep - (\omega_{l-k} - \omega_l))\Bigg|,
\end{align}
where $\hat{A}_p(0, 2\pi r \freqsep - (\omega_{l-k} - \omega_l))$ can be written as 
\begin{align}
\hat{A}_p(0, 2\pi r \freqsep - (\omega_{l-k} - \omega_l)) = \frac{e^{j (2\pi r \freqsep - (\omega_{l-k} - \omega_l)) \pri}-1}{j (2\pi r \freqsep - (\omega_{l-k} - \omega_l))}.
\end{align}
Since $(2\pi r \freqsep - (\omega_{l-k} - \omega_l))$ is always an integer multiple of $2\pi/\pri$, $e^{j (2\pi r \freqsep - (\omega_{l-k} - \omega_l)) \pri}$ always equals to $1$. We define $X_{k,r}(l) = \hat{A}_p(0, 2\pi r \freqsep - (\omega_{l-k} - \omega_l))/\pri$, which can be written as
\begin{align}\label{eq:Xkrl}
X_{k,r}(l) = \left\{
\begin{array}{ll} 1~~~~~~~~\omega_{l-k}-\omega_l = 2\pi r \freqsep\\
0~~~~~~~~\omega_{l-k}-\omega_l \neq 2\pi r \freqsep.
\end{array}\right.
\end{align}
Using \eqref{eq:Xkrl}, we re-write \eqref{eq:AF_FSK_grid2} as
\begin{align}\label{eq:AF_FSK_grid}
\Tilde{A}(k,r) = \Bigg| \frac{1}{\numslots}\sum_{l = k}^{\numslots - 1}X_{k,r}(l) \Bigg| , 
\end{align}
It can be observed from \eqref{eq:AF_FSK_grid} that the SL at $(t=k\pri, \omega = 2\pi r \freqsep)$ depends on how many frequency pairs differ by $2\pi r \freqsep$ at the $k$-th shift. In order to analyse the impact of frequency modulation on the SL statistically, we first study the probability distribution of $X_{k,r}(l)$. When $\numfreqs$-ary FSK modulation is used, the probability $\Pr[X_{k,r}(l)=1] = \Pr[\omega_{l-k}-\omega_l = 2\pi r \freqsep] = (\numfreqs - |r|)/\numfreqs^2$, indicating that $X_{k,r}(l)$ follows a Bernoulli distribution, i.e., 
\begin{align}
X_{k,r}(l) \sim \text{Bern} \left(\frac{\numfreqs - |r|}{\numfreqs^2}\right). 
\end{align}
Moreover, since the frequencies of different sub-pulses are mutually independent, $X_{k,r}(l), l \in \{k, \cdots, \numslots - 1\}$, whose values depend on whether $\omega_{l-k}-\omega_l = 2\pi r \freqsep$ as in \eqref{eq:Xkrl}, are independent and identically distributed (i.i.d). The independence between $X_{k,r}(l)$ and $X_{k,r}(l'), \forall l' \notin \{l, l-k, l+k\}$ is obvious since they depend on different pairs of frequencies. For $X_{k,r}(l)$ and $X_{k,r}(l-k)$, although both values depend on $\omega_{l-k}$, since $\omega_l$ and $\omega_{l-2k}$ are independent, it can be shown that there exists independence between $X_{k,r}(l)$ and $X_{k,r}(l-k)$, similar for $X_{k,r}(l)$ and $X_{k,r}(l+k)$. Therefore, the sum, $\sum_{l = k}^{\numslots - 1}X_{k,r}(l)$, follows a binomial distribution, i.e., 
\begin{align}
\sum_{l = k}^{\numslots - 1}X_{k,r}(l) \sim \text{B}\left(\numslots-k, \frac{\numfreqs-|r|}{\numfreqs^2} \right).   
\end{align}
Since $\Tilde{A}(k, r) = \sum_{l = k}^{\numslots - 1}X_{k,r}(l)/\numslots$, the probability mass function (PMF) of $\Tilde{A}(k, r)$ can be written as
\begin{align}  \label{eq:p_A_k_r}
    p_{\Tilde{A}(k,r)}\left(\frac{i}{\numslots}\right) &= {\numslots-k \choose i}\left(\frac{\numfreqs-|r|}{\numfreqs^2}\right)^i \left(\frac{\numfreqs^2-\numfreqs+|r|}{\numfreqs^2}\right)^{\numslots-k-i},\notag\\  i &\in \{0,\cdots, \numslots-k\}.
\end{align}
We can observe that the probability that $\Pr[\omega_{l-k}-\omega_l = 2\pi r \freqsep] =(\numfreqs-|r|)/\numfreqs^2$ decreases with $\numfreqs$, which means that the scaled binomial PMF in \eqref{eq:p_A_k_r} shifts toward the low random variable value region. Therefore, we conclude that the SL tends to decrease with $\numfreqs$. In addition, the number of Bernoulli trials, $\numslots-k$, decreases with the delay $k$, while $\Pr[\omega_{l-k}-\omega_l = 2\pi r \freqsep]$ decreases with the Doppler $r$, indicating that the SL further away from the origin tends to be smaller.



The AF grid point PSL is given as $\max_{(k,r) \in \mathcal{D}} \Tilde{A}(k,r)$. Since the SLs are neither mutually independent nor exchangeable, finding the distribution of their maximum is extremely challenging \cite{David03}. In order to analyse the statistics of the PSL, in this work we consider approximating the true distribution by assuming that the sidelobes are independent. This is motivated by the fact that although the sidelobes are not mutually independent, most pairwise correlations are $0$. It should be emphasised that we focus on only half of the delay-Doppler plane described by $(k,r) \in \mathcal{D}$ because of the symmetry of the AF. To illustrate the lack of correlation, we write $E\left[\Tilde{A}(k,r)\Tilde{A}(k',r')\right]$ as
\begin{align} 
E\left[\Tilde{A}(k,r)\Tilde{A}(k',r')\right] = \frac{1}{\numslots^2}\sum_{l=k}^{\numslots-1}\sum_{n=k'}^{\numslots-1}E\left[X_{k,r}(l)X_{k',r'}(n)\right]. 
\end{align}
The value of $E\left[\Tilde{A}(k,r)\Tilde{A}(k',r')\right]$ depends on $k$, $k'$, $r$ and $r'$. If $k \neq k'$, $X_{k,r}(l)$ is independent of $X_{k',r'}(n)$ since among all frequencies related to $X_{k,r}(l)$ and $X_{k',r'}(n)$, there is always at least one pair of independent frequencies. Thus, $E\left[X_{k,r}(l)X_{k',r'}(n)\right]$ can be written as $E\left[X_{k,r}(l)\right]E\left[X_{k',r'}(n)\right]$. If $k = k'$ and $r \neq r'$, however, $X_{k,r}(l)$ and $X_{k,r'}(n)$ are dependent when $l=n$, since both values depend on $\omega_{l-k}$ and $\omega_{l}$. More specifically, $E\left[X_{k,r}(l)\right]E\left[X_{k,r'}(l)\right] = 0$ since $\omega_{l-k}-\omega_l$ can never be equivalent to $2\pi r \freqsep$ and  $2\pi r' \freqsep$ simultaneously. Thus, $E\left[\Tilde{A}(k,r)\Tilde{A}(k',r')\right]$ can be re-expressed as 
\begin{align}\label{eq:E_Akr_Akprp}
&E\left[\Tilde{A}(k,r)\Tilde{A}(k',r')\right] \notag\\ 
&= \left\{
\begin{array}{ll} 
\frac{1}{\numslots^2}\sum_{l=k}^{\numslots-1}E\left[X_{k,r}(l)\right]\sum_{n=k'}^{\numslots-1}E\left[X_{k',r'}(n)\right]~k \neq k'\\
\frac{1}{\numslots^2}\underset{n\neq l}{\sum_{l=k}^{\numslots-1}\sum_{n=k}^{\numslots-1}}E\left[X_{k,r}(l)\right]E\left[X_{k,r'}(n)\right]~k = k', r \neq r'.
\end{array}\right.
\end{align}
The covariance between $\Tilde{A}(k,r)$ and $\Tilde{A}(k',r')$ can then be written as
\begin{align}\label{eq:cov_Akr_Akprp_1}
&E\left[\Tilde{A}(k,r)\Tilde{A}(k',r')\right] - E\left[\Tilde{A}(k,r)\right]E\left[\Tilde{A}(k',r')\right]
\notag\\ 
&= \left\{
\begin{array}{ll} 
0~~~~~~~~~~~~~~~~~~~~~~~~~~~~~~~~~~~~~~~~~k \neq k'\\
-\frac{1}{\numslots^2}\sum_{l=k}^{\numslots-1}E\left[X_{k,r}(l)\right]E\left[X_{k,r'}(l)\right]~k = k', r \neq r'.
\end{array}\right.
\end{align}
Using the Bernoulli mean, $E\left[X_{k,r}(l) \right] = (\numfreqs-|r|)/\numfreqs^2$, we re-write \eqref{eq:cov_Akr_Akprp_1} as
\begin{align}\label{eq:cov_Akr_Akprp}
&E\left[\Tilde{A}(k,r)\Tilde{A}(k',r')\right] - E\left[\Tilde{A}(k,r)\right]E\left[\Tilde{A}(k',r')\right]
\notag\\ 
&= \left\{
\begin{array}{ll} 
0~~~~~~~~~~~~~~~~~~~~~~~~~~~~~~~~~~k \neq k'\\
-(\numslots-k)(\numfreqs-|r|)^2/\left(\numslots^2\numfreqs^4\right)~k = k', r \neq r'.
\end{array}\right.
\end{align}
Dividing \eqref{eq:cov_Akr_Akprp} by the scaled binomial standard deviations, $\sqrt{\text{var}\left[\Tilde{A}(k,r) \right] \text{var}\left[\Tilde{A}(k',r') \right]} = (\numslots-k)(\numfreqs-|r|)(\numfreqs^2-\numfreqs+|r|)/\left(\numslots^2 \numfreqs^4\right)$, the correlation between $\Tilde{A}(k,r)$ and $\Tilde{A}(k',r')$ can be expressed as
\begin{align}\label{eq:corr_Akr_Akprp}
\rho_{\Tilde{A}(k,r)\Tilde{A}(k',r')} = \left\{
\begin{array}{ll} 
0~~~~~~~~~~~~~~~~~~k \neq k'\\
-\frac{\numfreqs-|r|}{\numfreqs^2-(\numfreqs-|r|)}~~~~~k = k', r \neq r'.
\end{array}\right.
\end{align}
We note from \eqref{eq:corr_Akr_Akprp} that the correlation between any two sidelobes is $0$, apart from those on the same delay axis. Furthermore, by noting that $|r|<\numfreqs$, the correlation between any two sidelobes on the same delay axis is negative. Its absolute value reaches the maximum, $1/(\numfreqs-1)$, at $|r|=0$, indicating that the correlation is small when $M$ is large.  

Motivated by the low or zero correlations shown in \eqref{eq:corr_Akr_Akprp}, we approximate the PSL distribution by assuming that the grid point sidelobes are mutually independent. The cumulative distribution function (CDF) of the maximum of independent random variables is simply the product of them, i.e.,
\begin{align} \label{eq:cdf_approx_ind}
F_{\max_{(k,r)\in \mathcal{D}}\Tilde{A}(k,r)}\left(\frac{i}{\numslots}\right) \approx &\prod_{(k,r) \in \mathcal{D}} F_{\Tilde{A}(k,r)}\left(\frac{i}{\numslots}\right), \notag\\  & 0 \leq i < \numslots,
\end{align}
where $F_{\Tilde{A}(k,r)}$ is the CDF of $\Tilde{A}(k,r)$, which can be written as
\begin{align}   
F_{\Tilde{A}(k,r)}\left(\frac{i}{\numslots}\right) = \sum_{q=0}^{\lfloor i\rfloor} {\numslots-k \choose q} & \left(\frac{\numfreqs-|r|}{\numfreqs^2}\right)^q \left(\frac{\numfreqs^2-\numfreqs+|r|}{\numfreqs^2}\right)^{\numslots-k-q},\notag\\ & 0 \leq i \leq \numslots-k.
\end{align}
Note that \eqref{eq:cdf_approx_ind} is the AF PSL distribution for the set of all FSK waveforms with one specific pair of $\numslots$ and $\numfreqs$.

We will later show via numerical examples that \eqref{eq:cdf_approx_ind} can accurately approximate the true CDF of the grid point PSL when $\numslots$ is large. Due to the accuracy, the approximation in \eqref{eq:cdf_approx_ind} can be used to reliably describe the actual statistical performance of the AF PSL. 
We can observe from both \eqref{eq:cdf_approx_ind} and later numerical examples that there is a variance in the AF PSL due to the use of independent FSK. For particular frequency sequences, the PSL can be much larger than well-designed radar waveforms such as Costas frequency codes a grid point PSL of $1/\numslots$ \cite{Levanon2004}. In other words, there exist FSK waveforms whose abilities to mitigate the effect of clutter presenting at particular delay-Doppler values are much worse compared to traditional radar waveforms. Therefore, it is important to suppress their PSLs. In the following section, we propose an optimisation-based method to minimise the PSL for each FSK waveform.

\section{PSL Minimisation with Optimal Phase Sequence} \label{eq:PSL_min}
In this work, we consider suppressing the grid point PSL of the AF by changing the initial phase of each sub-pulse. This is motivated by \cite{Zhou19}, where the zero-Doppler cut PSLs of PSK signals are suppressed using optimised phase perturbations. In contrast to \cite{Zhou19}, we consider the PSL among all grid points in the delay-Doppler plane. 
In addition, the phase change has different impacts on the communications performance of FSK and PSK. This will be discussed later in this section. 

\subsection{Phase Sequence Optimisation}
We first introduce the formulation of the optimisation problem. By changing the initial phases of all sub-pulses of the waveform in \eqref{eq:wf_FSK}, the new waveform can be expressed as   
\begin{align}
s_i(t) = \sqrt{\frac{1}{\numslots \pri}} \sum_{l=0}^{\numslots-1}\pulset \left(t- l\pri \right) \exp(j(\omega_l^i(t-l\pri)+\theta_l^i)), \label{eq:wf_ph}
\end{align}
where the index $i\in\{0,\cdots,\numfreqs^\numslots-1\}$ denotes the index of the waveform and $\omega_l^i$ and $\theta_l^i$ are the centre frequency and the initial phase of the $l$-th sub-pulse of the $i$-th waveform, respectively. The waveform index $i$ is required since the phase sequence, $\boldsymbol{\theta}^i = [\theta_0^i, \cdots, \theta_{\numslots-1}^i]^T$, is designed for the whole waveform but not a particular frequency symbol. The centre frequency of the $l$-th subpulse of the $i$-th waveform is given by the $l$-th digit of the waveform index $i$ written as a base-$M$ number of $L$ digits. For example, if $M=4$ and $L = 3$, the frequency sequence of the waveform with index $10$ will be $[0, 2\pi \times 2\freqsep, 2\pi \times 2\freqsep]$ as the $3$-digit base-$4$ representation of $10$ is $022$.

Substituting \eqref{eq:wf_ph} into \eqref{eq:cAF} and following similar steps as in \eqref{eq:cAF_FSK1}, \eqref{eq:cAF_FSK} and \eqref{eq:AF_FSK_grid1} to \eqref{eq:AF_FSK_grid}, we express the SL at the grid point $\tau = kT$ and $\omega = 2\pi r \Delta f$ as
\begin{align}\label{eq:AF_FSK_ph_grid}
\Tilde{A}_i(k,r;\boldsymbol{\theta}^i) = \left|\frac{1}{\numslots}\sum_{l = k}^{\numslots - 1}X_{k,r}(l) \exp(j(\theta^i_l-\theta^i_{l-k})) \right|,
\end{align}
Different from \eqref{eq:AF_FSK_grid}, the addends in \eqref{eq:AF_FSK_ph_grid} are complex-valued and thus the absolute value sign must be kept. 

Given the $i$-th FSK waveform, the corresponding phase sequence that minimises the PSL can be found by solving the following min-max optimisation problem
\begin{align} \label{eq:op1}
    \hat{\boldsymbol{\theta}^i} &= \argmin_{\boldsymbol{\theta}^i} \max_{(k,r) \in \mathcal{D}} \Tilde{A}_i(k,r;\boldsymbol{\theta}^i), \\
    \notag \text{s.t.,  } & 0 \leq \theta_l^i <2 \pi, l \in \{0,\cdots,\numslots-1\}.
\end{align}
Note that the optimal phase sequences for different FSK waveforms are different. Therefore, the optimisation problem in \eqref{eq:op1} should be solved for each waveform separately. 


We observe that the $\Tilde{A}(k,r)$'s lose their smoothness due to the absolute value sign in \eqref{eq:AF_FSK_ph_grid} and are not smooth functions of $\boldsymbol{\theta}$. To facilitate a better-behaved objective function, instead of using $\Tilde{A}(k,r)$, we adopt $\Tilde{A}^2(k,r)$ and re-formulate an equivalent problem to \eqref{eq:op1} as
\begin{align} \label{eq:op}
 \hat{\boldsymbol{\theta}^i} &= \argmin_{\boldsymbol{\theta}^i} \max_{(k,r) \in \mathcal{D}} \Tilde{A}^2_i(k,r;\boldsymbol{\theta}^i), \\
    \notag \text{s.t.,  } & 0 \leq \theta_l^i <2 \pi, l \in \{0,\cdots,\numslots-1\},
\end{align}
where 
\begin{align}\label{eq:AF2_FSK_ph_grid}
\Tilde{A}^2_i(k,r;\boldsymbol{\theta}^i) = \frac{1}{\numslots^2} & \sum_{l = k}^{\numslots - 1}\sum_{l' = k}^{\numslots - 1} X_{k,r}(l) X_{k,r}(l') 
\notag \\ &\times \exp(j((\theta_l^i-\theta_{l-k}^i)-(\theta_{l'}^i-\theta_{l'-k}^i))),
\end{align}
which is clearly first-order differentiable with respect to $\boldsymbol{\theta}^i$. 

While the design of min-max optimisation problem solvers has been widely studied and discussed \cite{Razaviyayn20}, it is outside the scope of this paper. In this work, we adopt the traditional and useful sequential quadratic programming (SQP) method \cite{Brayton79}, which can efficiently search for local minimum. Since the resulting local minimum is initial-value-dependent, for each waveform we perform the optimisation multiple times with different randomly generated initial values. We then select the local minimiser with the lowest objective function value as the final optimisation output. 

Since the number of FSK waveforms for given waveform parameters is large, i.e., $M^L$, the optimisation problem in \eqref{eq:op} has to be repeated many times, resulting in a huge computation cost even with the efficient SQP method. One way of reducing the cost is by noting that not all waveforms require optimisation-based PSL 
suppression. There are $M$ FSK waveforms in which the frequencies of all sub-pulses are the same. Existing phase codes that ensure low AF PSLs can be applied to these waveforms \cite{Levanon2004}, indicating that optimisation is not required. 
Nevertheless, the remaining $(M^L-M)$ waveforms still require PSL suppression. We can also decrease the dimensionality of the problem by fixing $\theta_0$ to $0$ and looking for $[\theta_1,\cdots,\theta_{\numslots-1}]$ that minimises the objective function \cite{Zhou19}. This works since the objective function value of \eqref{eq:AF2_FSK_ph_grid} remains the same if all initial phases are changed by the same amount, $\theta_0$. Moreover, due to the symmetry of the AF, we consider only half of the delay-Doppler plane, which has already been adopted in the definition of $\mathcal{D}$. The design of advanced techniques for reducing the computation cost is considered a future extension of the existing work.


\subsection{Impact on Communications Performance}

In this section, we focus on the impact of PSL suppression on the communications performance, in particular, the data rate and the communications error probability. For each FSK waveform, a single optimal phase sequence is selected, thus for a given $\numslots$ and $\numfreqs$, the number of waveforms remains unchanged. Therefore, the optimisation process does not affect the data rate, whose value is $(\log_2{\numfreqs})/\pri$ bits/sec.

\subsubsection{Maximum likelihood detection before PSL suppression}
In order to understand the impact of the phase sequence on the communications error probability, let us consider a system model similar to that in \cite{Han2023_freqpermPSK}, where the transmitter has a single antenna and the communications receiver has $\numantas$ antennas. The complex envelope of a single sub-pulse received at the communications receiver can be expressed as
\begin{align} \label{eq:comm_rcvd_nophase}
    \boldsymbol{r}(t) = \sqrt{\frac{1}{\numslots \pri}} \boldsymbol{h} p(t) \exp(j 2\pi \Tilde{m} \freqsep t) + \boldsymbol{n}(t),
\end{align}
where $\boldsymbol{h} \in \mathbb{C}^{\numantas \times 1}$ is the baseband complex channel fading vector, $2\pi \Tilde{m} \freqsep$ is the centre frequency of the transmitted sub-pulse and $\boldsymbol{n}(t)\in \mathbb{C}^{\numantas \times 1}$ is the complex additive white Gaussian noise (AWGN) with the power spectrum density (PSD) $N_0$. Note that we ignore the sub-pulse index $l$ since the data modulation is independent of the sub-pulse. We make the following assumptions: $\boldsymbol{h}$ is independent of $\boldsymbol{n}(t)$ and has a slow fading model in which the channel gain vector $\boldsymbol{h}$ remains unchanged within a coherence interval longer than the waveform duration $\numslots \pri$. We further assume that the communications receiver has perfect knowledge of $\boldsymbol{h}$, obtained by sending a set of training waveforms at the beginning of the coherence interval. The symbol-by-symbol maximum likelihood (ML) FSK receiver first correlates the received sub-pulse with all $\numfreqs$ basis functions, with the $m$-th is the sub-pulse whose centre frequency is $2\pi m \freqsep$ as \cite[Chapter 4]{Proakis08}
\begin{align} \label{eq:comm_corr_nophase1}
    y_{m} = &\int_0^{\pri} \sqrt{\frac{1}{\numslots \pri}}\exp(-j 2\pi m \freqsep t) \boldsymbol{h}^H \boldsymbol{r}(t)  d t, \notag \\ &m \in \{0,\cdots, \numfreqs-1\}.
\end{align}
The $m$-th correlation, conditioned on the transmission of the sub-pulse with the $k$-th frequency, can be expressed as
\begin{align} \label{eq:comm_corr_nophase2}
y_m = \mu_m^k + n_m 
\end{align}
where 
\begin{align} \label{eq:comm_mean_nophase}
\mu_m^k = \left\{
\begin{array}{ll} 
\sqrt{1/\numslots}\|\boldsymbol{h}\|_2^2 ~~~~~ m = k\\
0~~~~~~~~~~~~~~~~ m \neq k,
\end{array}\right.
\end{align}
and $n_{m} = \sqrt{1/(\numslots \pri)} \int_0^{\pri} \boldsymbol{h}^H\boldsymbol{n}(t) \exp(-j 2\pi m \freqsep t) d t$ is the noise component.
Note that $n_{m}, m \in \{0, \cdots,\numfreqs-1\}$ are independent and identically distributed (i.i.d) $\sim \mathcal{CN}\left(0,\|h\|^2_2 N_0/ \numslots\right)$ since the basis functions are orthogonal and contain the same amount of energy. The $k$-th likelihood function is the joint probability density function (PDF) of $y_{m}$'s, $m \in \{0,\cdots,\numfreqs-1\}$, conditioned on the transmission of the sub-pulse with the centre frequency $2\pi k \freqsep$, which can be expressed as
\begin{align} \label{eq:comm_likelihood_nophase1}
    &f\left(y_{0},\cdots,y_{\numfreqs-1} | k \right) \notag  \\ 
    = &\frac{1}{\left(\pi\|\boldsymbol{h}\|_2^2N_0/\numslots\right)^{\numfreqs}} \exp\left(-\frac{\sum_{m=0}^{\numfreqs-1}\left|y_m-\mu^k_m\right|^2}{\|\boldsymbol{h}\|_2^2N_0/\numslots}\right).
\end{align}
The maximum likelihood detector can then detect the information embedded in the sub-pulse by maximising $f\left(y_{0},\cdots,y_{\numfreqs-1} | k \right)$, or equivalently, 
\begin{align} \label{eq:comm_rcver}
\hat{m} = \argmax_{k \in \{0, 1, \cdots, \numfreqs-1\}} \psi_k, 
\end{align}
where
\begin{align}  \label{eq:comm_likelihood_nophase}
\psi_k = \re (y_k),
\end{align}
with $\re(\cdot)$ denoting the operation of taking the real part. The derivation of the equivalent detector is straightforward by substituting  \eqref{eq:comm_mean_nophase} into \eqref{eq:comm_likelihood_nophase1} and removing terms in \eqref{eq:comm_likelihood_nophase1} that are constant with respect to the optimisation.

\subsubsection{Maximum likelihood detection after PSL suppression}
After performing the PSL suppression, the complex envelope of the received signal when the $\Tilde{i}$-th waveform is transmitted can be expressed as
\begin{align} \label{eq:comm_rcvd_phase}
    \boldsymbol{r}(t) = \boldsymbol{h} s_{\Tilde{i}}(t) + \boldsymbol{n}(t).
\end{align}
Since the PSL suppression process can be performed in advance for all waveforms, we can assume that the initial phase terms are known at the communications receiver. However, the optimised phase sequence is associated with each waveform and not each FSK symbol. Therefore, unlike the simple symbol-by-symbol detector in \eqref{eq:comm_rcver}, the ML detector has to take the whole waveform into consideration. We start from correlating the received signal with all $\numfreqs^{\numslots}$ basis functions, each of them denoting the sub-pulse in a specific time slot with a specific centre frequency as
\begin{align} \label{eq:comm_corr_withphase1}
    y_{m,l} = &\int_{l \pri}^{(l+1)\pri} \sqrt{\frac{1}{\numslots \pri}}\exp(-j 2\pi m \freqsep t) \boldsymbol{h}^H \boldsymbol{r}(t)  d t, \notag \\ &m \in \{0,\cdots, \numfreqs-1\}, l \in \{0,\cdots, \numslots-1\}.
\end{align}
Conditioned on the transmission of $s_i(t)$, $y_{m,l}$ can be expressed as
\begin{align} \label{eq:comm_corr_withphase2}
    y_{m,l} = \mu^i_{m,l} + n_{m,l}
\end{align}
where 
\begin{align} \label{eq:comm_mean_withphase}
\mu^i_{m,l} = \left\{
\begin{array}{ll} 
\sqrt{1/\numslots}\|\boldsymbol{h}\|_2^2 \exp(j\theta_l^i) ~~~~~ m = k_l^i\\
0~~~~~~~~~~~~~~~~~~~~~~~~~~ m \neq k_l^i,
\end{array}\right.
\end{align}
with $k_l^i$ denoting the frequency index of the $l$-th sub-pulse of $s_i(t)$. The i.i.d noise components, $n_{m,l}=\sqrt{1/(\numslots \pri)} \int_{l \pri}^{(l+1)\pri} \boldsymbol{h}^H \boldsymbol{n}(t) \exp(-j 2\pi m \freqsep t) d t \sim \mathcal{CN}\left(0,\|h\|^2_2 N_0/ \numslots\right), m \in \{0,\cdots, \numfreqs-1\}, l \in \{0,\cdots, \numslots-1\}$. The joint PDF of $y_{m,l}$'s, conditioned on the transmission of $s_i(t)$, can be expressed as
\begin{align} \label{eq:comm_likelihood_withphase1}
    &f\left(y_{0,0}, \cdots, y_{\numfreqs-1,\numslots-1}|\boldsymbol{s}_i(t)\right) \notag \\
    = &\frac{1}{\left(\pi\|\boldsymbol{h}\|_2^2N_0/\numslots\right)^{\numslots\numfreqs}} \exp\left(-\frac{\sum_{l=0}^{\numslots-1}\sum_{m=0}^{\numfreqs-1}\left|y_{m,l}-\mu^i_{m,l}\right|^2}{\|\boldsymbol{h}\|_2^2N_0/\numslots}\right).
\end{align}
The ML detector for the $l$-th received sub-pulse has a similar formulation as \eqref{eq:comm_rcver} with a different likelihood function, which is the joint PDF of $y_{m,l}$'s conditioned on the event that the $l$-th sub-pulse of the transmitted waveform has a centre frequency $2\pi k \freqsep$. It can be derived using the law of total probability as 
\begin{align} \label{eq:comm_likelihood_withphase2}
 f\left(y_{0,0}, \cdots, y_{\numfreqs-1,\numslots-1}|k \right) = \sum_{i=0}^{\numfreqs^\numslots} \Pr\left(\boldsymbol{s}_i(t)|\omega_l^i = 2\pi k \freqsep\right)& \notag \\
 \times f\left(y_{0,0}, \cdots, y_{\numfreqs-1,\numslots-1}|\boldsymbol{s}_i(t), \omega_l^i = 2\pi k \freqsep \right)&.
\end{align}
Noting that $\Pr\left(\boldsymbol{s}_i(t)|\omega_l^i = 2\pi k \freqsep\right) = 1/\numfreqs^{\numslots-1}$ when the $l$-th frequency of $s_i(t)$ is $2\pi k\freqsep$ and $0$ otherwise, \eqref{eq:comm_likelihood_withphase2} can be re-expressed as
\begin{align} \label{eq:comm_likelihood_withphase3}
 &f\left(y_{0,0}, \cdots, y_{\numfreqs-1,\numslots-1}|k \right) \\ \notag = &\frac{1}{\numfreqs^{\numslots-1}} \sum_{i \in \mathcal{I}_{l,k}} f\left(y_{0,0}, \cdots, y_{\numfreqs-1,\numslots-1}|\boldsymbol{s}_i(t)\right),
\end{align}
where $\mathcal{I}_{l,k} \subset \{0,\cdots, \numfreqs^\numslots-1\}$ denotes the set of indices of waveforms whose $l$-th subpulse has a centre frequency $2\pi k \freqsep$. Note that $|\mathcal{I}_{l,k}| = \numfreqs^{\numslots-1}$. An equivalent ML detector can be formulated by substituting \eqref{eq:comm_likelihood_withphase1} into \eqref{eq:comm_likelihood_withphase3} and removing terms that are constant with respect to the optimisation, resulting in the objective function as 
\begin{align} \label{eq:comm_likelihood_withphase}
\xi_{k} = \sum_{i \in \mathcal{I}_{l,k}} \exp\left(-\frac{\sum_{l=0}^{\numslots-1}\sum_{m=0}^{\numfreqs-1}\left|y_{m,l}-\mu^i_{m,l}\right|^2}{\|\boldsymbol{h}\|_2^2N_0/\numslots}\right).
\end{align}

\subsubsection{Sub-optimal detection after PSL suppression}
Although both likelihood functions in \eqref{eq:comm_likelihood_nophase} and \eqref{eq:comm_likelihood_withphase} detect one received symbol, \eqref{eq:comm_likelihood_withphase} uses the whole received waveform with $\numslots$ sub-pulses. Therefore, the computation of $\xi_{k}$ requires an extremely large lookup table (LUT) with all $\numfreqs^\numslots$ waveforms stored at the receiver, which is significantly less efficient compared to \eqref{eq:comm_likelihood_nophase} with an $\numfreqs$-element LUT. In addition, the computation of $\xi_{k}$ is more complicated than $\psi_k$ since the calculation of exponential functions and the summation of selected terms are required.

The problem of receiver complexity can be mitigated by noting that the initial phase does not carry information. Thus, instead of detecting the waveform index, we treat the initial phase as an unknown phase noise and consider a similar symbol-by-symbol receiver as in \eqref{eq:comm_rcver} with a different objective function, $\phi_k$, which can be expressed as \cite[Chapter 7]{richards14}
\begin{align}  \label{eq:comm_noncoherent_withphase}
\phi_k = \left|y_k \right|.
\end{align} 
Note that $\psi_k$ and $\phi_k$ are the real part and the magnitude of the correlation $y_k$, respectively. The detectors with $\psi_k$ and $\phi_k$ as likelihood functions are exactly the coherent and the non-coherent FSK detectors, respectively. The corresponding error rates for different channel models have been analysed in detail in \cite{Argyriou00}. Note that the error probabilities for FSK with and without the initial phase terms are the same when the non-coherent detector is used in both situations. This can be proved by analysing the distribution of \eqref{eq:comm_noncoherent_withphase}. Suppose the transmitted subpulse is $\sqrt{1/(\numslots \pri)} p(t) \exp (j (2\pi \Tilde{m}\freqsep t + \Tilde{\theta}) )$, the correlation $\phi_k$ can be expressed as 
\begin{align} \label{eq:corr_phase1}
\phi_k  = \left\{
\begin{array}{ll} 
\left|\sqrt{(1/\numslots)}\|\boldsymbol{h}\|_2^2 \exp(j\Tilde{\theta}) + n_k\right|~~~~~  k = \Tilde{m}\\
 \left| n_k\right|~~~~~~~~~~~~~~~~~~~~~~~~~~~~~~~~ k  \neq \Tilde{m},
\end{array}\right.
\end{align}
which can be further expressed as
\begin{align} \label{eq:corr_phase}
\phi_m  = \left\{
\begin{array}{ll} 
\left|\sqrt{(1/\numslots)}\|\boldsymbol{h}\|_2^2 + \exp(-j\Tilde{\theta}) n_k\right|~~~~~ k  = \Tilde{m}\\
 \left| n_k\right|~~~~~~~~~~~~~~~~~~~~~~~~~~~~~~~~~ k  \neq \Tilde{m}.
\end{array}\right.
\end{align}
Since $n_k$ is circularly symmetric Gaussian distributed, $n_k$ and \\$\exp(-j\Tilde{\theta}) n_k$ are identically distributed. Note that $n_k$ is exactly the noise component when the initial phase term is not introduced, indicating that the distribution of \eqref{eq:comm_noncoherent_withphase} and the corresponding error rate is irrelevant to the initial phases. Thus, the error performance of non-coherent detection for standard FSK \cite{Argyriou00} can be applied to the case after PSL suppression.

\section{Numerical Examples} \label{sec:num_FSK_global}  





\begin{figure*}[t]
\centering
\includegraphics[width=18cm]{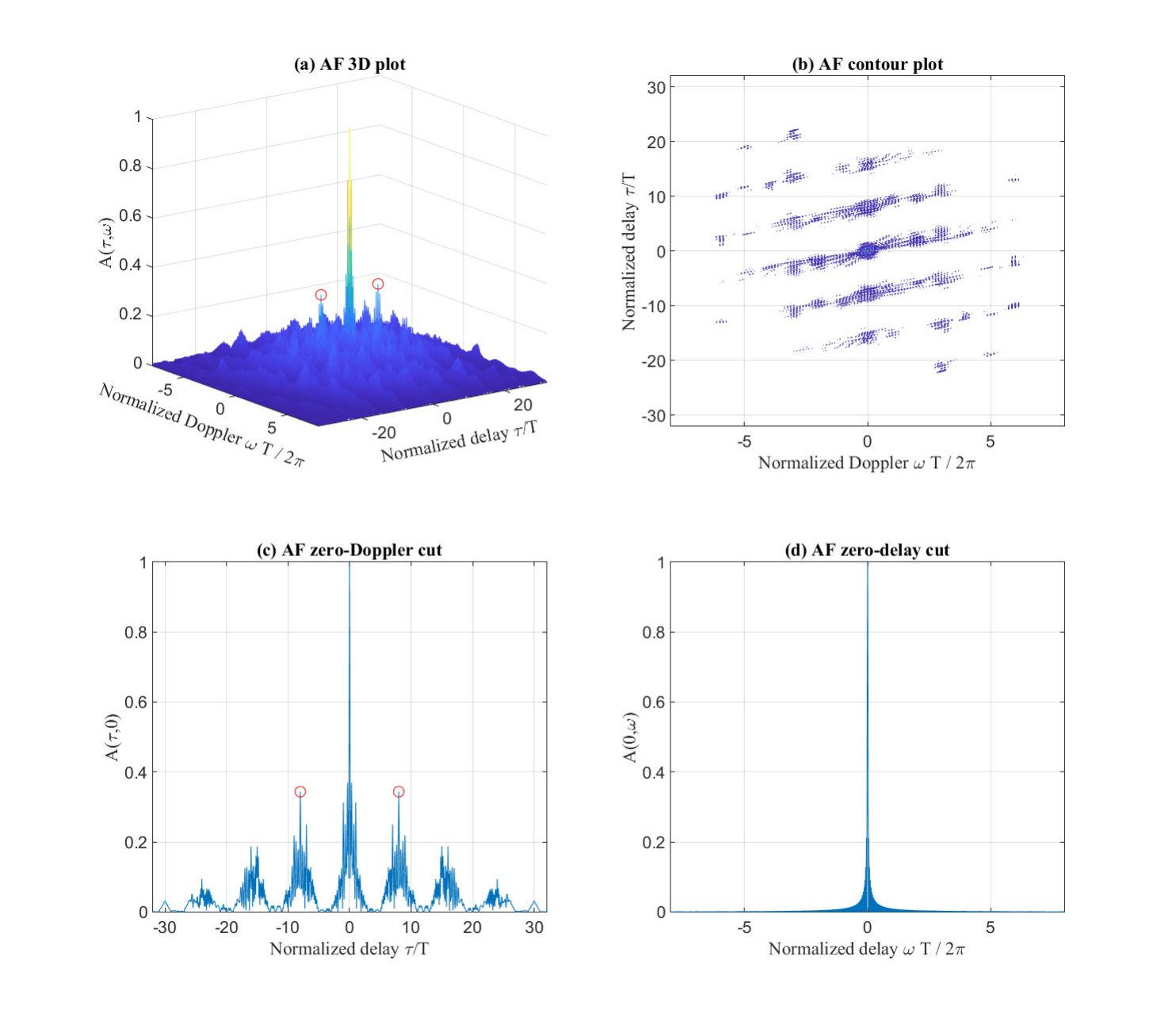}
\caption{The (a) three-dimensional (3D) surface plot, (b) contour plot, (c) zero-Doppler cut and (d) zero-delay cut of the AF of a specific realisation of FSK waveform with $\numslots=32$ and $\numfreqs=8$.
}\label{fig:AF_random}
\end{figure*}

Figure \ref{fig:AF_random} plots the AF and its one-dimensional cuts for one specific realisation of the FSK waveform with $\numslots=32$ and $\numfreqs=8$ before PSL suppression. 
Sub-figure \ref{fig:AF_random}(a) plots the three-dimensional AF surface. We can observe that although SLs at the majority of the delay-Doppler plane are low, the grid point PSL of this AF is non-ideal with a value of $11/\numslots$ compared to the $1/\numslots$ PSL of the Costas-coded waveform, as is shown by the two red circles. Sub-figure \ref{fig:AF_random}(b) plots the two-dimensional contour of the AF in sub-figure \ref{fig:AF_random}(a), which more clearly shows the structure of the AF. Sub-figures \ref{fig:AF_random}(c) and \ref{fig:AF_random}(d) provide more insight into the AF by plotting the zero-Doppler cut and the zero-delay cut which represent the delay and the Doppler estimation ability of the waveform, respectively. We observe from sub-figure \ref{fig:AF_random}(c) that the PSL appears at $\tau = \pm 8/\pri, \omega = 0$. Sub-figure \ref{fig:AF_random}(d) presents the AF zero-delay cut of all FSK waveforms with $\numslots=32$ as it does not change with the selection of the frequency sequence.

Next, we investigate if the grid point PSL is a useful lower bound on the local maxima PSL and if \eqref{eq:cdf_approx_ind} well approximates the grid point PSL distribution. Both comparisons require evaluating the difference between distributions, at least one of which is empirical. In this work, we consider the Wasserstein $1$-distance, which can be calculated as \cite[Proposition 1]{Aaditya17}
\begin{align} \label{eq:was_dis1}
    W_1 = \int_{-\infty}^{\infty} \left| F_1^{-1}(x) - F_2^{-1}(x)\right| dx,
\end{align}
where $F_1(x)$ and $F_2(x)$ are the CDFs of the two distributions. Using the obvious fact that 
\begin{align}
    \int_{-\infty}^{\infty} \left| F_1(x) - F_2(x)\right| dx = \int_{0}^{1} \left| F_1^{-1}(y) - F_2^{-1}(y)\right| dy,
\end{align}
we re-write \eqref{eq:was_dis1} as
\begin{align} \label{eq:was_dis}
    W_1 = \int_{-\infty}^{\infty} \left| F_1(x) - F_2(x)\right| dx.
\end{align}
Note that \eqref{eq:was_dis} is the $L_1$ norm of the difference between two CDFs, which captures the total area between them. Compared to other commonly used measures such as the two-sample Kolmogorov-Smirnov test statistic \cite{Massey51}, this measure considers both deviations in the probabilities and the PSL values and is, therefore, a more complete characterisation of the difference between two CDFs. 

\begin{figure}[t]
\centerline{\includegraphics[width=8cm,height=6.5cm]{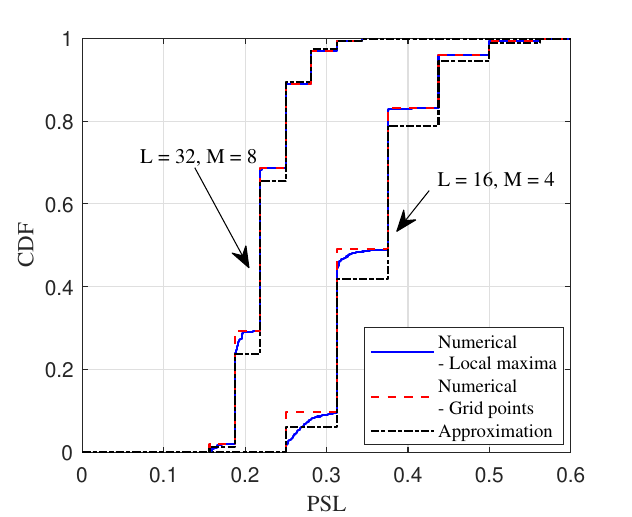}}
    \caption{The empirical CDF of the local maxima PSL, the empirical CDF of the grid point PSL and the approximation in \eqref{eq:cdf_approx_ind} for $\numslots = 16, \numfreqs = 4$ and $\numslots = 32, \numfreqs = 8$.}\label{fig:PSL_CDF_L16M4_L32M8}
\end{figure}

Figure \ref{fig:PSL_CDF_L16M4_L32M8} compares the empirical CDFs of the local maxima PSL (blue solid curve) and the grid point PSL (red dashed curve) for $\numslots = 16, \numfreqs = 4$ and $\numslots = 32, \numfreqs = 8$. The Wasserstein $1$-distances between the CDFs for the two sets of parameters are $0.0023$ and $0.0005$, respectively. In addition, we note that the horizontal difference between the two CDFs is upper-bounded by $1/\numslots$, indicating that the grid point PSL lower bounds the local maxima PSL well. Figure \ref{fig:PSL_CDF_L16M4_L32M8} also compares the empirical CDF of the grid point PSL and the approximate CDF (black dotted-dashed curve) in \eqref{eq:cdf_approx_ind} computed based on the independence assumption. The Wasserstein $1$-distances for the two sets of parameters are $0.0223$ and $0.0032$, respectively, indicating that the approximation in \eqref{eq:cdf_approx_ind} is accurate. In order to investigate the impact of waveform parameters $\numslots$ and $\numfreqs$ on the Wasserstein $1$-distance between the grid point PSL CDF and its approximation, Table \ref{tab:was_dis_LM} lists its values for $\numslots,\numfreqs \in \{4,8,16,32,64,128,256\}$. We observe that the Wasserstein $1$-distance decreases with $\numslots$, either with a fixed $\numfreqs$ or a fixed ratio between $\numslots$ and $\numfreqs$. This indicates that the approximation proposed in \eqref{eq:cdf_approx_ind} has high accuracy when the number of sub-pulses $\numslots$ is large. 

\begin{table}[t] 
\caption{The Wasserstein $1$-distance between the CDF of the grid point PSL and its approximation for $\numslots,\numfreqs \in \{4,8,16,32,64,128,256\}$} 
\vspace{-1.5\baselineskip}
\begin{center}
\renewcommand{\arraystretch}{1.1}
\begin{tabular}{m{0.9cm}<{\centering}| m{0.6cm}<{\centering} m{0.6cm}<{\centering} m{0.6cm}<{\centering} m{0.6cm}<{\centering} m{0.6cm}<{\centering} m{0.6cm}<{\centering} m{0.7cm}<{\centering}}
\hline\hline
\diagbox{\numfreqs}{\numslots} & $4$ & $8$ & $16$ & $32$ & $64$ & $128$ & $256$ \\ [1ex]
\hline
$4$ & $0.0609$ &	$0.0142$ & $0.0223$ & $0.0046$ & $0.0041$ & $0.0017$ & $0.0007$ \\[0.5ex]
$8$ & $0.0447$ &	$0.0238$ & $0.0094$ & $0.0032$ & $0.0045$ & $0.0017$ & $0.0011$ \\[0.5ex]
$16$ & $0.0337$ & $0.0189$  & $0.0104$ & $0.0080$ & $0.0019$ & $0.0007$ & $0.0004$ \\[0.5ex]
$32$ & $0.0192$ & $0.0409$  & $0.0042$ & $0.0034$ & $0.0017$ & $0.0010$ & $0.0004$ \\[0.5ex]
$64$ & $0.0119$ & $0.0371$  & $0.0070$ & $0.0011$ & $0.0014$ & $0.0007$ & $0.0003$ \\[0.5ex]
$128$ & $0.0075$ & $0.0205$ & $0.0091$ & $0.0015$ & $0.0010$ & $0.0004$ & $0.0002$ \\[0.5ex]
$256$ & $0.0036$ & $0.0095$ & $0.0206$ & $0.0024$ & $0.0005$ & $0.0002$ & $0.0002$ \\[0.5ex]
\hline \hline
\end{tabular}
\label{tab:was_dis_LM}  
\end{center}
\end{table}

\begin{figure}[t]
    \centerline{\includegraphics[width=8cm]{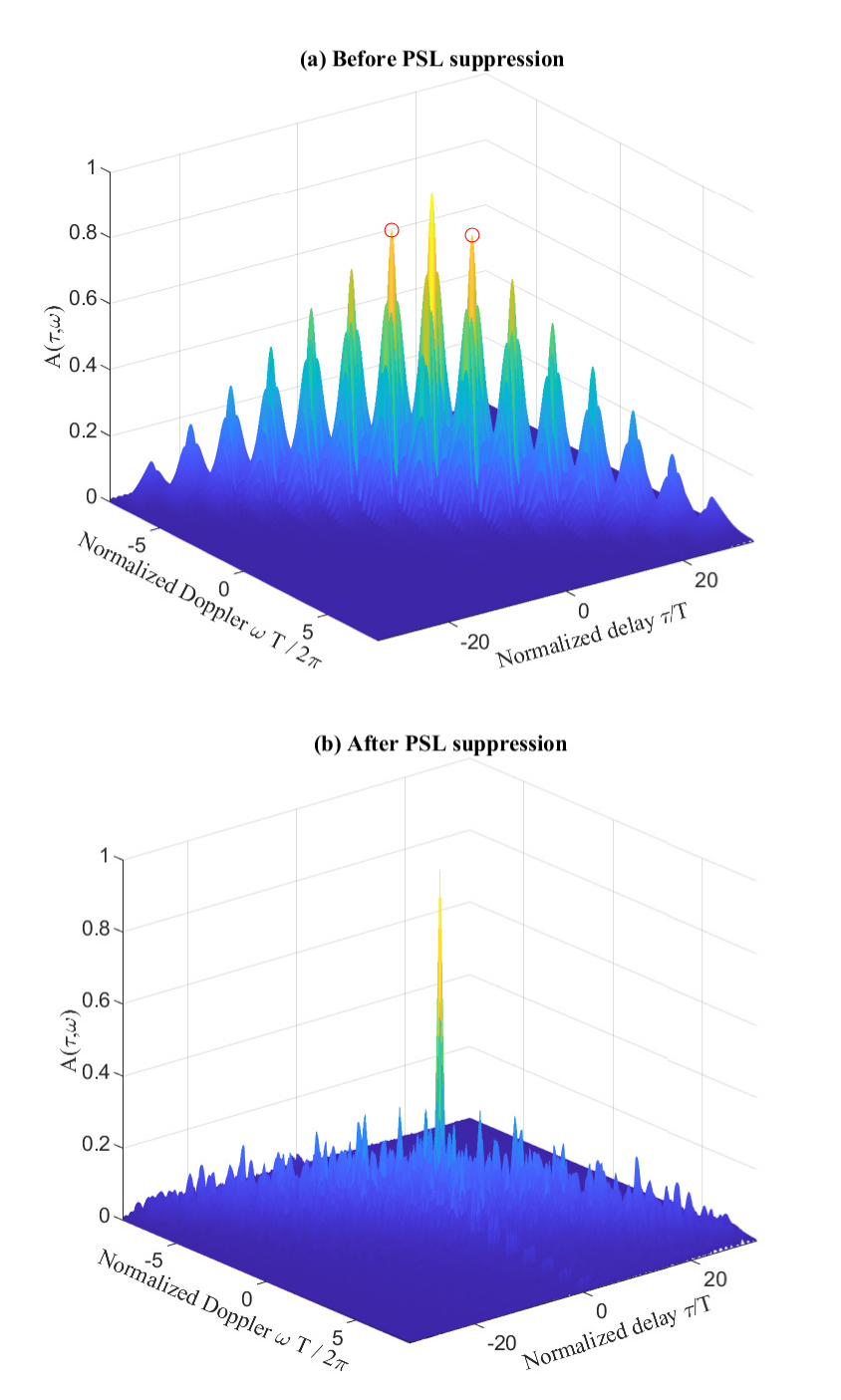}}
    \caption{The AFs of an example FSK waveform with $\numslots = 32$ and $\numfreqs = 8$ (a) before and (b) after the PSL suppression with $\numslots = 32$ and $\numfreqs = 8$.}\label{fig:af_L32M8}
\end{figure}

Next, we present numerical examples to describe the drop in the grid point PSL when the optimal phase sequence is introduced. Figure \ref{fig:af_L32M8} compares the AFs of an FSK waveform with $\numslots = 32$ and $\numfreqs = 8$ before and after the PSL suppression. The particular selection of the frequency sequence 
leads to a big ridge in the AF, resulting in a huge PSL of $28/\numslots = 0.875$ which is shown by the two red circles. After introducing the optimised phase sequence, the PSL is significantly suppressed to around $1.7/\numslots = 0.053$, indicating that the proposed method is effective.  

To investigate the effect of the PSL suppression for different waveform parameters $\numslots$ and $\numfreqs$, Table \ref{tab:PSL_op} and Table \ref{tab:PSL_drop} list the average PSL value and the average drop in PSL after the suppression for $\numslots \in \{4,8,16,32,64\}$ and $\numfreqs \in \{2,4,8\}$, respectively. We observe from Table \ref{tab:PSL_op} that the average PSL after suppression is at the level of $\mathcal{O}(1/\numslots)$, indicating that the proposed method leads to a PSL close to the well-designed radar waveforms such as Costas codes \cite{Levanon2004}. In addition, we observe from Table \ref{tab:PSL_drop} that the PSL drop decreases with $\numfreqs$. This is mainly because the PSL before suppression is generally low when $\numfreqs$ is large as the probability of matching between frequency pairs, $(\numfreqs-|r|)/\numfreqs^2$, decreases with $\numfreqs$; while the PSL after suppression decreases insignificantly with $\numfreqs$. 


\begin{table}[t] 
\caption{The average PSL after the suppression for $\numslots \in \{4,8,16,32,64\}$ and $\numfreqs \in \{2,4,8\}$.}  
\vspace{-1.5\baselineskip}
\begin{center}
\renewcommand{\arraystretch}{1.1}
\begin{tabular}{{m{1cm}<{\centering}| m{0.675cm}<{\centering} m{0.675cm}<{\centering} m{0.675cm}<{\centering} m{0.675cm}<{\centering} m{0.8cm}<{\centering}}}
\hline\hline
\diagbox{\numfreqs}{\numslots}  & $4$ & $8$ & $16$ & $32$ & $64$ \\ [1ex]
\hline
$2$ & $0.2500$ & $0.1328$	& $0.0883$	& $0.0592$	& $0.0418$ \\[0.5ex]
$4$ & $0.2500$ & $0.1273$	& $0.0822$	& $0.0577$	& $0.0389$ \\[0.5ex]
$8$ & $0.2500$ & $0.1256$	& $0.0710$	& $0.0557$	& $0.0350$	\\[0.5ex]
\hline \hline
\end{tabular}
\label{tab:PSL_op}  
\end{center}
\end{table}

\begin{table}[t] 
\caption{The average PSL drop after the suppression for $\numslots \in \{4,8,16,32,64\}$ and $\numfreqs \in \{2,4,8\}$.}  
\vspace{-1.5\baselineskip}
\begin{center}
\renewcommand{\arraystretch}{1.1}
\begin{tabular}{{m{1cm}<{\centering}| m{0.675cm}<{\centering} m{0.675cm}<{\centering} m{0.675cm}<{\centering} m{0.675cm}<{\centering} m{0.8cm}<{\centering}}}
\hline\hline
\diagbox{\numfreqs}{\numslots}  & $4$ & $8$ & $16$ & $32$ & $64$ \\ [1ex]
\hline
$2$ & $0.1875$ & $0.3936$	& $0.4649$	& $0.5004$	& $0.5164$ \\[0.5ex]
$4$ & $0.0996$ & $0.2394$	& $0.2695$	& $0.2760$	& $0.2778$ \\[0.5ex]
$8$ & $0.0518$	& $0.1644$	& $0.1819$	& $0.1650$	& $0.1603$	\\[0.5ex]
\hline \hline
\end{tabular}
\label{tab:PSL_drop}  
\end{center}
\end{table}

\begin{figure}[t]
    \centerline{\includegraphics[width=8cm,height=6.5cm]{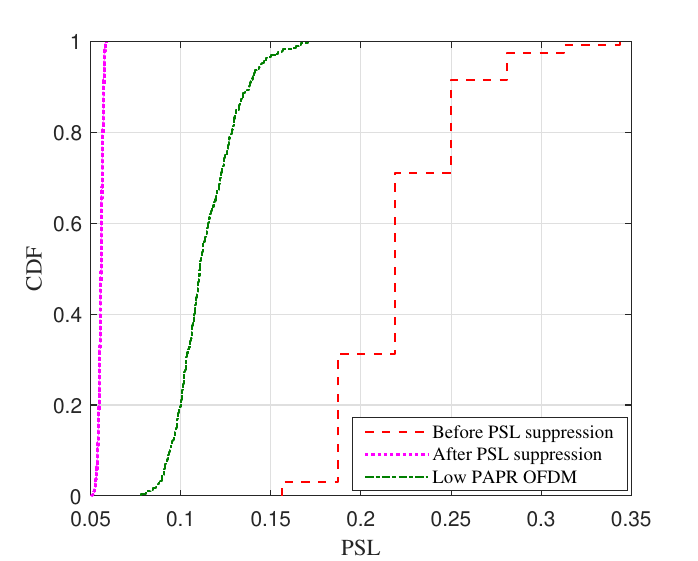}}
    \caption{The empirical CDFs of the grid point PSL before and after PSL suppression, and the low PAPR OFDM waveform \cite{Piyush23} for $\numslots = 32$ and $\numfreqs = 8$. The DBR for the OFDM scheme is $3/8$ and the modulation scheme is BPSK.}\label{fig:PSL_CDF_suppression_L32M8}
\end{figure}

Figure \ref{fig:PSL_CDF_suppression_L32M8} plots the empirical CDFs of the grid point PSL before (red dashed line) and after (magenta dotted line) the suppression for $\numslots = 32$ and $\numfreqs = 8$. When the optimised phase sequence is introduced, the PSL is decreased significantly to around $1.8/\numslots = 0.056$, with a small variation among different waveforms. 
Given that the optimisation procedure does not sacrifice the data rate, our proposed scheme has the same data rate as a traditional communications scheme, while its AF PSL performance is comparable to well-designed radar waveforms.  
In addition, we compare our proposed scheme with a low PAPR OFDM-based JCR waveform \cite{Piyush23} (green dash-dotted line). Unlike our proposed waveform which has a unit baseband PAPR value due to its single-tone subpulse nature, \cite{Piyush23}  sacrifices the data rate to address the issue of the high PAPR of OFDM. More specifically, for each OFDM symbol a portion of subcarriers are designed to minimise the PAPR while the remaining subcarriers are used for data transmission. Generally speaking, the resulting PAPR decreases with the data-bandwidth ratio (DBR), i.e., the ratio between the number of communications subcarriers and the total number of subcarriers. We denote the number of symbols and the number of subcarriers of the OFDM-based waveform by $\numslots$ and $\numfreqs$, respectively. We choose $\numslots = 32$, $\numfreqs = 8$, DBR $=3/8$ and the modulation scheme of BPSK for the low PAPR OFDM waveform. These parameters result in the same waveform time duration, bandwidth and data rate, making the comparison relatively fair. Note that in \cite{Piyush23}, only BPSK and QPSK modulation schemes are considered due to the requirement of low zero-Doppler SLs and the effectiveness of the PAPR minimisation. The grid point PSL of the OFDM waveform is around $0.114$ with a larger variation, indicating that our proposed scheme performs better in terms of the radar global accuracy. 
\textcolor{red}{Will cite \cite{Piyush23} in the figure after the reference list is finalised}

Table \ref{tab:FSK_vs_OFDM} provides a thorough comparison between our proposed waveform and \cite{Piyush23} in terms of the average grid point PSL, average PAPR and the data rate for different waveform parameters. Note that the data rates of the proposed scheme and the low PAPR OFDM scheme are $\log_2\numfreqs/T$ and $\text{DBR} \times (\numfreqs \log_2 \ddot{M})/T$, respectively, where $\ddot{M}$ denotes the modulation order used in OFDM. For all low PAPR OFDM waveforms, the stopping criterion for the PAPR minimisation algorithm \cite[Algorithm 1]{Piyush23} is either a maximum of $800$ iterations or a convergence threshold of $10^{-6}$. Although the OFDM-based scheme has the flexibility to trade-off between the data rate and the radar performance by tuning the DBR, our proposed waveform outperforms \cite{Piyush23} in terms of both the average PSL and the average PAPR under the same data rate. More specifically, the average PSL of our proposed scheme is less than half of that of the OFDM-based scheme. In addition, the PAPR value of the proposed waveform is always equivalent to $1$, which is lower than the average PAPR of the OFDM-based waveform. 
\begin{table}[t] 
\caption{Comparison between the proposed FSK with optimised phases scheme and the low PAPR OFDM scheme \cite{Piyush23}.}  
\vspace{-1.5\baselineskip}
\begin{center}
\renewcommand{\arraystretch}{1.1}
\begin{tabular}{{m{2.7cm}<{\centering}| m{1.35cm}<{\centering} m{1.35cm}<{\centering} m{1.5cm}<{\centering}}}
\hline\hline
Signalling scheme & Average PSL & Average PAPR & Data rate (bits/sec)\\ [1ex]
\hline
Proposed, $\numslots=8$, $\numfreqs=4$ & $0.1278$ & $1$ & $2/\pri$	 \\[0.5ex]
\hline
\cite{Piyush23} , DBR $= 1/2$, BPSK, $\numslots=8$, $\numfreqs=4$ & $0.2615$ & $1.0684$	& $2/\pri$\\[0.5ex]
\hline
\cite{Piyush23} , DBR $= 1/2$, QPSK, $\numslots=8$, $\numfreqs=4$ & $0.2635$ & $1.1220$	& $4/\pri$	 \\[0.5ex]
\hline
Proposed, $\numslots=32$, $\numfreqs=8$ & $0.0557$ & $1$ & $3/\pri$	\\[0.5ex]
\hline
\cite{Piyush23} , DBR $= 3/8$, BPSK, $\numslots=32$, $\numfreqs=8$ & $0.1140$ & $1.1231$	& $3/\pri$\\[0.5ex]
\hline
\cite{Piyush23} , DBR $= 3/8$, QPSK, $\numslots=32$, $\numfreqs=8$ & $0.1246$ & $1.1286$	& $6/\pri$	 \\[0.5ex]
\hline \hline
\end{tabular}
\label{tab:FSK_vs_OFDM}  
\end{center}
\end{table}

\begin{figure}[t]
    \centerline{\includegraphics[width=8cm,height=6.5cm]{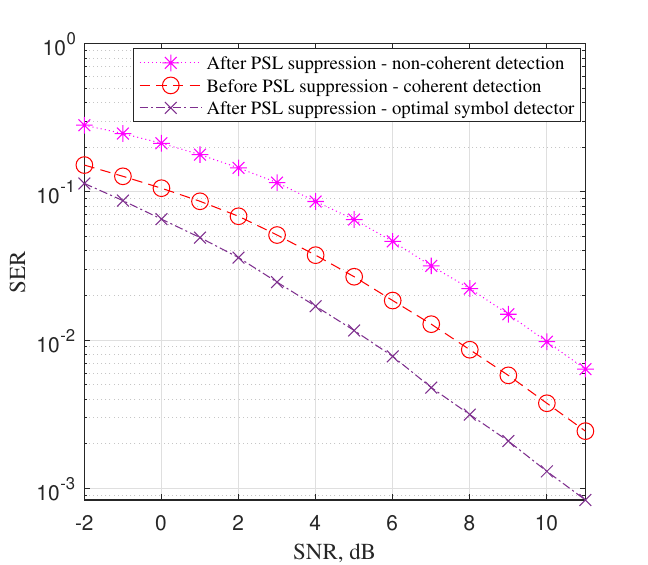}}
    \caption{SER versus SNR before and after PSL suppression for $\numslots=3$, $\numfreqs = 2$ under Rician fading channels with $2$ receive antennas and $K$ factor equivalent to $1$.}\label{fig:ser_rician_L3M2N2K1}
\end{figure}

\begin{figure}[t]
\centerline{\includegraphics[width=8cm,height=6.5cm]{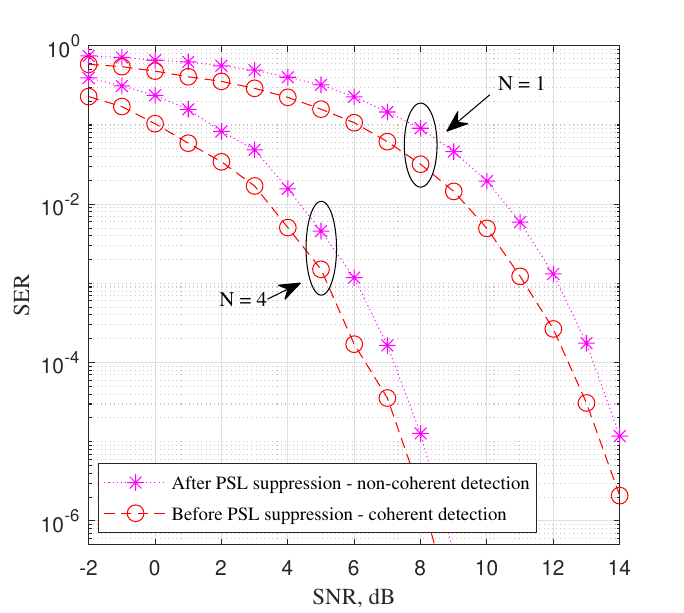}}
    \caption{SER versus SNR before and after PSL suppression for $\numslots=32$, $\numfreqs = 8$ under AWGN channels and $N \in \{1,4\}$ receive antennas.}\label{fig:ser_awgn_M8N1_4}
\end{figure}

Figure \ref{fig:ser_rician_L3M2N2K1} and Figure \ref{fig:ser_awgn_M8N1_4} plot the symbol error probability (SER) versus average signal-to-noise ratio (SNR) curves before and after the PSL suppression is performed. Note that the SNR value is the ratio between the sub-pulse energy and the PSD, averaged over different channel fading realisations. 
Figure \ref{fig:ser_rician_L3M2N2K1} considers Rician fading channels with $N=2$ receive antennas and $K$ factor equivalent to $1$. The Rician $K$ factor is chosen such that both the line-of-sight path and the scatter have sufficient components. Small waveform parameters, i.e., $\numslots=3$ and $\numfreqs=2$ are considered to show the performance of the inefficient optimal symbol detector after the PSL suppression. We observe that the optimal symbol detector after the PSL suppression outperforms that before the suppression by $2.40$ dB SNR in the low SER region since the minimum distance between waveforms is increased after the initial phase terms are introduced. The SNR gap between the curves for the efficient non-coherent detector after the suppression and the coherent one before the suppression decreases with the decrease in the SER. In the low SER region, the SNR gap is around $2.23$ dB.
Figure \ref{fig:ser_awgn_M8N1_4} considers waveform parameters $\numslots=32$ and $\numfreqs=8$ and AWGN channels with $N \in \{1,4\}$ receive antennas. The optimal symbol detection after PSL suppression is not performed due to its inefficiency when waveform parameters are large. Again, the SNR gap between the curves decreases with the decrease in the SER. In the low SER region, the SNR gaps are around $0.69$ dB and $0.62$ dB, respectively. Therefore, for non-small waveform parameters, using non-coherent detection after PSL suppression has a small increase in the error probability compared to using coherent detection before PSL suppression while keeping the receiver complexity unchanged. Considering the PSL and its drop in Table \ref{tab:PSL_op} and Table \ref{tab:PSL_op}, we conclude that our proposed method effectively suppresses the AF PSL while having an insignificant impact on the communications receiver performance. 




\section{Conclusion and Future Extensions} \label{sec:conclusion}
Motivated by the unit PAPR property and the ideal radar sensing performance of frequency-coded radar waveforms, in this work we consider the FSK-based waveform for JCR. We mainly focus on analysing the impact of frequency modulation on the radar performance by deriving the probability distributions of the AF SLs. In addition, an approximation of the AF PSL distribution is provided. Numerical results show that the approximation is highly accurate, especially for waveforms with a large number of sub-pulses. The derived probability distributions indicate that uncontrollable AF SLs and PSLs are introduced by the independent frequency modulation. Therefore, we formulate a min-max optimisation problem to design the initial phases of waveform sub-pulses for the purpose of minimising the AF PSL. Numerical results show that the existing SQP-based min-max optimisation solver can effectively reduce the AF PSL to $1/\numslots$, which is close to the well-designed Costas-coded radar waveform. In addition, we show that introducing initial phases has an insignificant impact on the communications performance in terms of the data rate, the receiver complexity and the error probability. Compared to a state-of-the-art low PAPR OFDM-based JCR benchmark waveform, the proposed waveform has an average AF PSL of less than half of the benchmark and a better average PAPR under the condition of the same data rate. 

Since the PSL suppression process is not real-time and has to be performed only once for each waveform, the efficiency of the optimisation algorithm is not that critical. However, the time required to find the optimal phase sequences for all waveforms with large $\numslots$ and $\numfreqs$ is still too long, though some methods that reduce the computation cost are considered. The main reasons include the number of waveforms to be optimised $\numfreqs^\numslots-\numfreqs$, the number of PSL values to be considered in each min-max optimisation problem $|\mathcal{D}|=2\numslots\numfreqs+\numslots+\numfreqs$ and the dimensionality of the problems $\numslots-1$. 
Therefore, advanced techniques for reducing the computation cost are required and considered a future extension of the existing work.

\ifCLASSOPTIONcaptionsoff
  \newpage
\fi



%



\bibliographystyle{IEEEtran}


\end{document}